\documentclass{XrU2005}

\title{Galactic Corona or Local Group Intergalactic Medium?}
\author{Rik J.~Williams}
\author{Smita Mathur}
\affil{Department of Astronomy, The Ohio State University, 140 W. 18th Ave., Columbus, OH 43210 USA}
\author{Fabrizio Nicastro}
\affil{Harvard--Smithsonian Center for Astrophysics, Cambridge, MA  USA}

\newcommand{\kms}{\,km\,s$^{-1}$}

\newcommand{\chandra}{\emph{Chandra}}
\newcommand{\fuse}{\emph{FUSE}}
\newcommand{\hi}{H\,{\sc i}}
\newcommand{\novii}{N_{\rm OVII}}
\newcommand{\novi}{N_{\rm OVI}}

\newcommand{\ovilv}{O\,{\sc vi}$_{\rm LV}$}
\newcommand{\ovi}{O\,{\sc vi}}
\newcommand{\ovii}{O\,{\sc vii}}
\newcommand{\oviii}{O\,{\sc viii}}
\newcommand{\nvii}{N\,{\sc vii}}

\usepackage{graphicx}

\begin{document}


\maketitle

\begin{abstract}
Cosmological hydrodynamic simulations predict that the low redshift universe 
comprises a web of warm--hot intergalactic gas and galaxies, with groups of 
galaxies and clusters forming at dense  knots in these filaments. Our own 
Galaxy 
being no exception is also expected to be  surrounded by the warm--hot 
intergalactic medium, filling the Local Group. Some theoretical  models 
also predict the existence of a hot Galactic corona.  With X-ray and 
FUV  observations of extragalactic sources, we can probe the warm--hot 
gas through absorption lines  of highly ionized elements. Indeed, \chandra, 
\emph{XMM} and \fuse\ observations have detected $z=0$ absorption lines toward 
many sightlines. The debate that has emerged is over the  interpretation 
of these observations: are the $z=0$ absorption systems from the halo of our  
Galaxy or from the extended Local Group environment? This has important 
implications for  our understanding of the mass of the Local Group, the 
physical conditions in the intergalactic  medium, the structure of the 
Galaxy and galaxy formation in general. We will present the  current status 
of the debate and discuss our ongoing observing program aimed at  
understanding the $z=0$ absorption systems, with an emphasis on the 
high quality \chandra\ spectra of the Mrk~421 and Mrk~279 sightlines.   
\end{abstract}

\section{Introduction}
The intergalactic medium (IGM) is expected to contain most of the baryonic
matter in the universe, a tenuous filamentary ``web'' of gas bridging
the gaps between collapsed objects such as galaxies and clusters.
At high redshifts ($z>2$) this web appears in quasar spectra as a multitude
of Lyman alpha forest absorption lines.  In the nearby universe, on the
other hand, hydrodynamic simulations show that most of the IGM has
been shock--heated to a warm--hot (WHIM) phase with temperatures of
$\sim 10^6$\,K \citep{cen99,dave01}.  At these temperatures scant neutral 
hydrogen
remains, and the IGM is thus best detected through absorption lines
from highly ionized metals \citep{hellsten98}, particularly \ovi, \ovii, and
\oviii.  Recent \chandra\ observations have indeed confirmed that this 
low--redshift WHIM exists and comprises a baryon content consistent with 
expectations \citep[Nicastro et al., in preparation][]{nicastro05a,nicastro05b}.

Just as other galaxies are expected to form in the densest ``knots''
of the cosmic web, we also expect to see WHIM adjacent to, perhaps 
surrounding, our own Milky Way.   Indeed, X-ray spectra of several quasars
show likely $z=0$ \ovii\ absorption.  The upper limit on the \ovii\ 
emission toward Mrk 421 found by \citet{rasmussen03} indicates that this
absorption system probably has an extremely low density and is thus
likely to be extragalactic.  Additionally, other nearby low-- and 
high--ionization components may be associated with either the WHIM itself
or may represent gas from the WHIM that has cooled and is now in the
process of accreting onto the Galaxy; for example, the high--velocity
\ovi\ absorption seen with \fuse\ along many quasar lines of sight
\citep{wakker03} and neutral hydrogen high--velocity clouds (\hi\ HVCs).

\section{The Debate: WHIM or Corona?}
The origin of the local \ovii\ absorption, and in particular its relation
to the observed \ovi, is still to a large degree unknown.  There is some
evidence that these ions could originate in a warm--hot Galactic corona:
for example, likely \ovii\ absorption has been detected within 50
kpc of the Galaxy by \citet{wang05}.  The observed deflection and
stripping of the Magellanic clouds also lends credence to the existence
of a low--density corona.  Lower--ionization absorption, such as that
from Si\,{\sc iv} and C\,{\sc iv}, is also seen at the same velocities
as some \ovi\ HVCs, indicating that these \ovi\ clouds, at least, may
have lower temperatures and higher densities than expected from the WHIM.

However, there are also reasons to believe this absorption is tracing
extended WHIM gas.  High column densities of \ovi, \ovii, and \oviii\ 
from the local IGM are predicted by simulations to lie in certain 
directions \citep{kravtsov02}.  Furthermore, the mean velocity vector of the
\ovi\ HVCs is highest in the local standard of rest and lowest in the
Local Group rest frame, indicating that their origin could indeed
be extragalactic \citep{nicastro03}.  The presence of \ovii\ between
the Galaxy and Large Magellanic Cloud does not rule out an extragalactic
origin for the absorption in some directions since neither the Galactic
absorption or WHIM is necessarily homogeneous; additionally, the WHIM
is known to be homogeneous and consist of a variety of temperature and
density phases, so some lower--ionization lines may be expected as well.

Thus, the questions -- how are the local X-ray and UV absorption
components related to each other, which are of Galactic origin, and
which arise in the local WHIM?-- are still unanswered.  
The answers to these questions have profound implications for both
studies of galaxy formation and cosmology.
We are now undertaking a program to determine the distribution and
properties of this local hot gas and its ties to lower--ionization
components, with a focus on new and archival \chandra\ and \fuse\ data. 
Here we present the first results of this study, an analysis of the
particularly high--quality \chandra\ and \fuse\ spectra of the 
bright AGN Mrk~421 and Mrk~279.

\section{The Mrk~421 Sightline}
A full discussion of the Mrk~421 \chandra\ and FUSE spectra, and the
analysis thereof, can be found in \citet{williams05}; the following is
a summary of the main results.

\subsection{Observations and measurements}
The bright $z=0.03$ blazar 
Mkn~421 was observed during two exceptionally high
outburst phases for 100 ks each as part of our \chandra --AO4
observing program:
one at $f_{\rm 0.5-2 keV}=1.2\times 10^{-9}$\,erg\,s$^{-1}$\,cm$^{-2}$
with the Low Energy Transmission Grating (LETG) combined with the
ACIS-S array, and another
at $f_{\rm 0.5-2 keV}=0.8\times 10^{-9}$\,erg\,s$^{-1}$\,cm$^{-2}$
with the HRC-S array and LETG.  Each of these observations
contains $\sim 2500$ counts per resolution element at 21.6\,\AA.  Additionally,
another short observation of Mkn~421 was taken with HRC/LETG
(29 May 2004), providing another 170 counts per resolution element.  These
three spectra were combined over the 10--60\,\AA\ range to improve the
signal--to--noise ratio (S/N$\sim 55$ at 21\,\AA\ with 0.0125\,\AA\ binning).
The final coadded spectrum of Mkn~421 is one of the best ever taken
with \chandra: it contains
over $10^6$ total counts with $\sim 6000$ counts per resolution element at
21.6\,\AA, providing a $3\sigma$ detection threshold of
$W_\lambda\sim 2$\,m\AA\
($N_{\rm OVII}=8\times 10^{14}\rm{cm}^{-2}$ for an unsaturated line).

Using the CIAO fitting package Sherpa we initially modeled the continuum
of Mkn~421 as a simple power law with Galactic foreground absorption,
excluding the 48--57\,\AA\ HRC chip gap region.  Metal abundances for
the Galactic gas were then artificially adjusted to provide a better fit
around the \ion{O}{1} and \ion{C}{1} K--edges near 23\,\AA\ and 43\,\AA\
respectively.  This is \emph{not} intended to represent actual
changes to the absorber composition, but rather to correct uncertainties
in the instrument calibration.  
After this fit there were still some systematic uncertainties in
the best--fit continuum model; these were corrected with
broad (${\rm FWHM}=0.15-5$\,\AA) Gaussian emission and absorption components
until the modeled continuum appeared to match the data upon inspection.
Indeed, the residuals of the spectrum to the final continuum model have a
nearly Gaussian distribution, with a negative tail indicating the presence
of narrow absorption lines \citep[see][Figure 8]{nicastro05a}.
We searched for narrow, unresolved (FWHM$<0.05$\,\AA) absorption lines 
at known C, N, O, and Ne transition wavelengths and used Gaussians
to measure the equivalent widths (or upper limits thereupon) for all
lines found.  All in all, equivalent widths for 9 significantly detected
absorption lines were measured (including
\ovii\ K$\alpha$, $\beta$, and $\gamma$) and 4 upper limits calculated.

\begin{figure}
\centering
\includegraphics[width=0.8\linewidth]{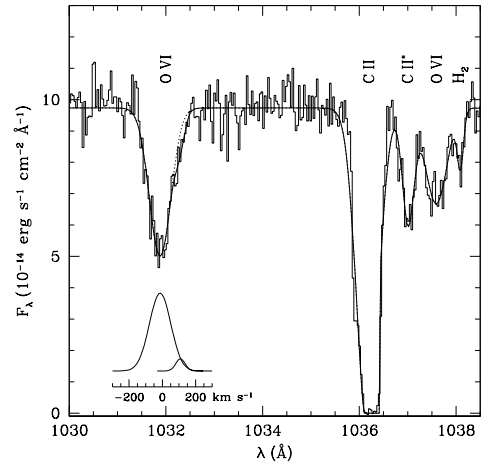}
\caption{FUSE spectrum of Mrk~421 near the \ovi\ doublet.  The 1032\,\AA\ 
line is well--fit by a strong absorber at $v\sim 0$\kms\ and a much weaker
component at $v\sim 100$\kms\ (inset).\label{fig_mkn421fuse}}
\end{figure}

Mrk~421 was also observed for a total of 84.6\,ks with FUSE, providing
a signal--to--noise ratio of 17 near the \ovi\ wavelength once
all spectra are combined and binned to $\sim 10$\kms.  This spectrum
shows strong, broad \ovi\ absorption at $v \sim 0$, most likely originating
in the Galactic thick disk, as well as a possible \ovi\ HVC at
$v\sim 110$\kms\ (Figure~\ref{fig_mkn421fuse}).

\subsection{Doppler parameters}
To convert the measured equivalent widths to ionic column densities,
we calculated curves of growth for each absorption line
over a grid of Doppler parameters ($b=10-100$\kms) and column densities
($\log N_H/{\rm cm}^{-2} = 12.0-18.0$), assuming a Voigt line profile.  
Since the X-ray lines are unresolved, $b$ cannot be measured directly.
It can, however, be inferred from the relative strengths of the
three measured \ovii\ K--series lines.  
These line ratios by themselves are insufficient to determine the physical
state of the \ovii--absorbing medium since $b$ and $\novii$ are
degenerate: the K$\alpha$ line saturation could be due to high column
density, low $b$, or a combination of both.  However, given an absorption 
line with a measured equivalent width and known
oscillator strength, the inferred column density as a function of the
Doppler parameter can be calculated.  The measured equivalent width 
(and errors) for
each transition thus defines a region in the $\novii-b$ plane.
Since the actual value of $\novii$ is fixed, $b$ and $\novii$ can be
determined by the region over which the contours ``overlap;''  i.e.
the range of Doppler parameters for which the different transitions
provide consistent $\novii$ measurements.

\begin{figure}
\centering
\includegraphics[width=0.8\linewidth]{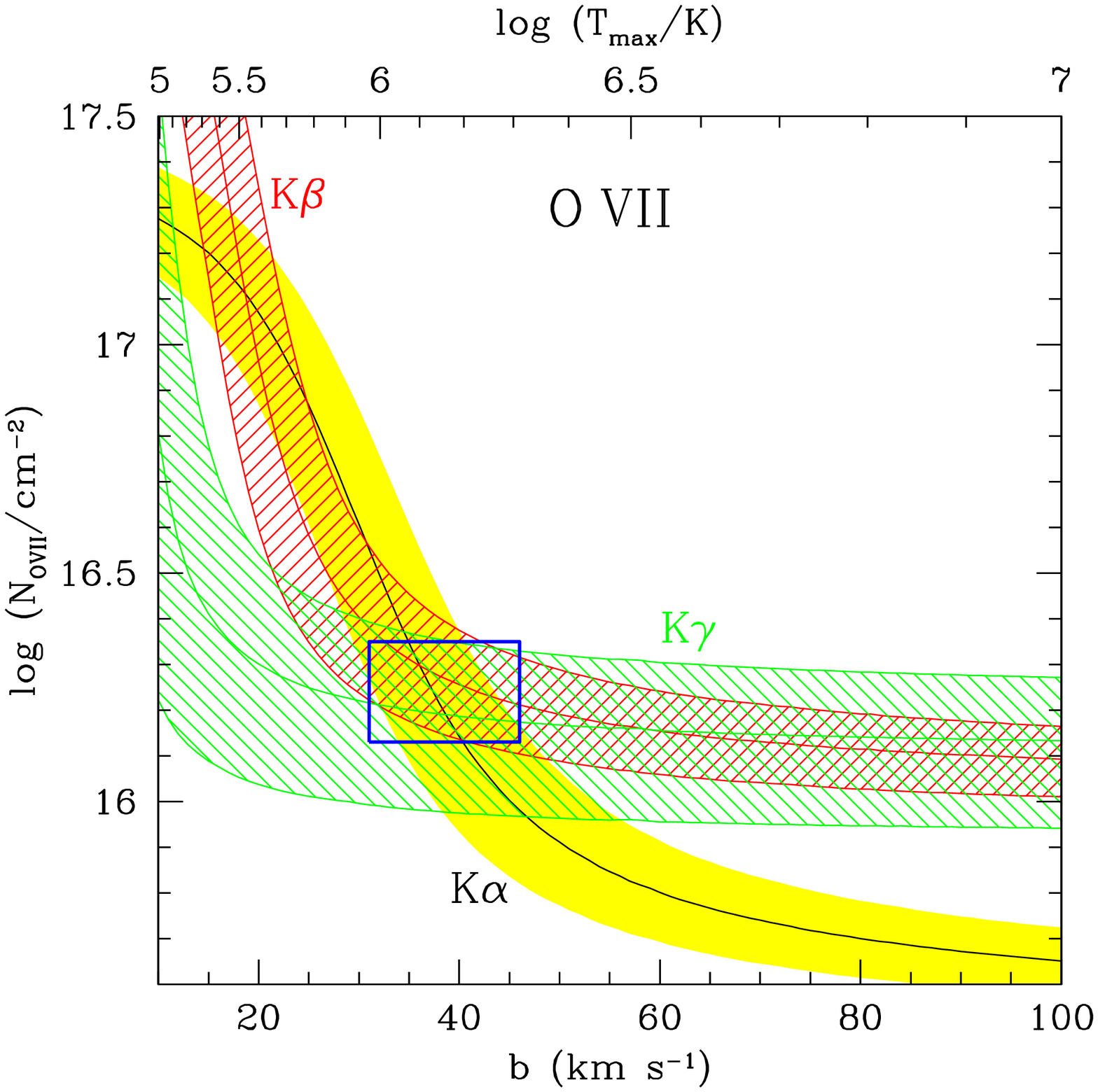}
\includegraphics[width=0.8\linewidth]{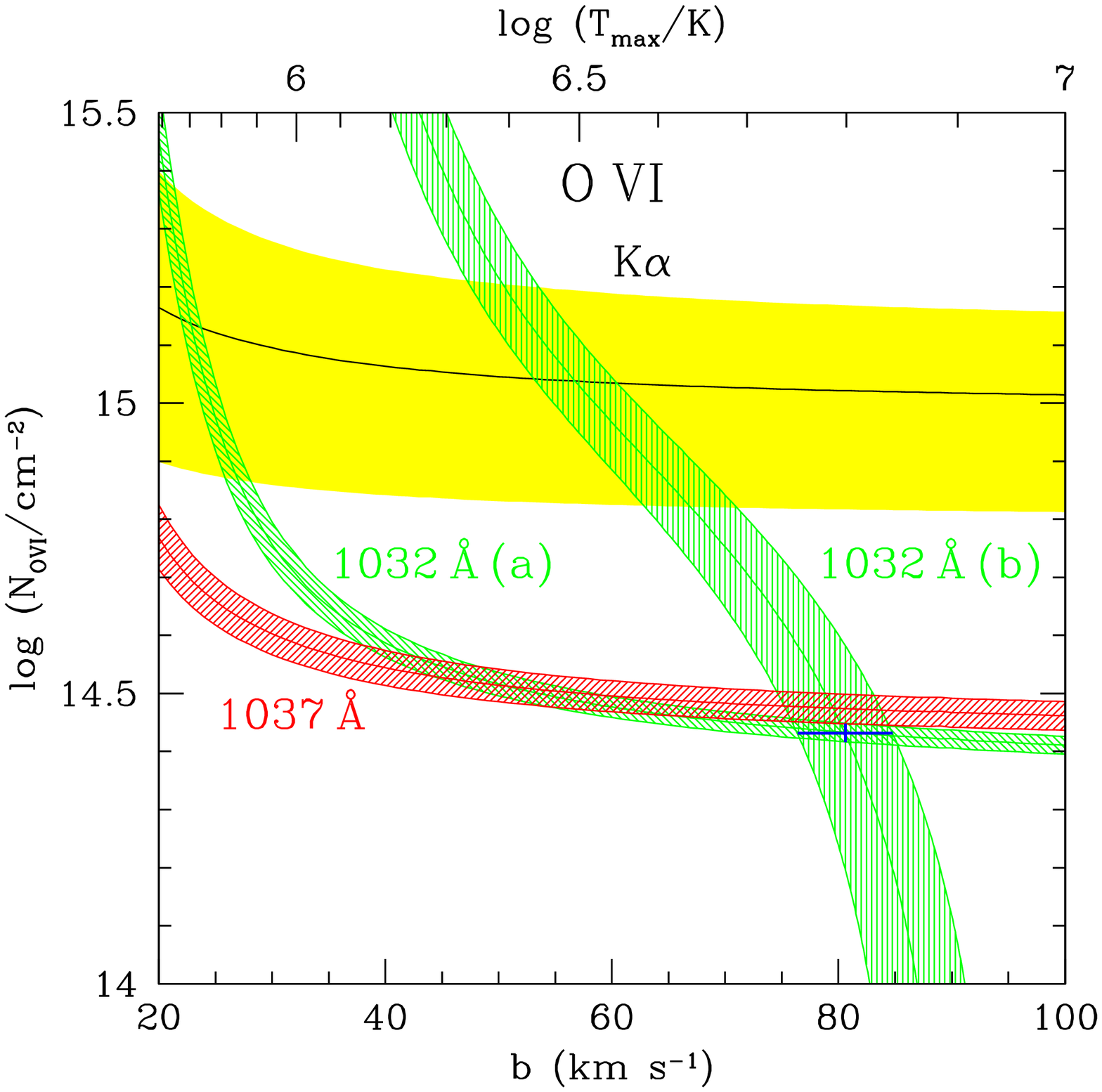}
\caption{Mrk 421 column density and Doppler parameter diagnostics for 
the \ovii\ (top) and \ovi\ (bottom) absorption, with each transition
labeled.  In the \ovi\ panel, the ``a'' and ``b'' curves are derived from
the 1032\,\AA\ absorption equivalent width and FWHM, respectively. 
\label{fig_nb_mkn421}}
\end{figure}

Figure~\ref{fig_nb_mkn421} (top panel) shows such $1\sigma$ contours for the 
three measured
\ovii\ transitions.  The K$\alpha$ and K$\beta$ tracks appear consistent
at the $2\sigma$ level for $13<b<55$\kms,
while the overlap between the K$\alpha$ and K$\gamma$ tracks provides
approximate 2$\sigma$ limits of $24<b<76$\kms.  
We thus assume a $2\sigma$ range
of $24<b<55$\kms.  It should be noted that Figure~\ref{fig_nb_mkn421} also
shows some overlap between the $K_\alpha$ and $K_\gamma$ at $b\le 12$\kms; 
however, this solution is unlikely given the 
lower limit provided by the $K_\beta$ line.  Moreover, $b=12$\kms\ implies
a maximum temperature (assuming purely thermal motion) of 
$T_{\rm max}=1.3\times 10^5$\,K; such a low temperature is unlikely to 
produce the observed strong high--ionization lines.

A similar analysis is not as effective when applied to the strong \ovilv\ 
UV doublet (from the thick disk), since these lines are only slightly 
saturated.  The \ovilv\ 1032\,\AA\ line is fully 
resolved by \fuse\ 
and relatively unblended, so its Doppler parameter can be estimated much more
accurately using the measured line width and strength.  In an unsaturated
absorption line, ${\rm FWHM}=2(\ln 2)^{1/2}b$; however, the measured FWHM
increases if the line is saturated.  We compensated for this by 
calculating Voigt profile FWHMs on a grid of $\novi$ and $b$,
and determining the region consistent with the \ovilv\ 1032\,\AA\ FWHM 
measurement of $152\pm 7$\kms.

When the FWHM--derived contour is overlaid on the $\novi-b$ contour inferred
from the equivalent width measurement of the LV--\ovi\ 1032\,\AA\ line, 
the two regions overlap nearly orthogonally 
(Figure~\ref{fig_nb_mkn421}, bottom panel) leading to a constraint of 
$b$(\ovilv)$=80.6\pm 4.2$\kms.  This is more than $3\sigma$ higher than
the Doppler parameter calculated for the \ovii\ absorption, indicating
that the \ovii\ and thick--disk \ovilv\ cannot arise in the same gaseous
phase.  Also, at no value
of the Doppler parameter do the 1032\,\AA, 1037\,\AA, and \ovi\ K$\alpha$
lines all produce a consistent $\novi$ measurement; in fact, the
\ovi K$\alpha$ column density is a factor of $\sim 4$ higher than that
inferred from the UV data.  It is thus possible that the \ovi\ UV
transition is being suppressed, perhaps by atomic physics effects, in the 
absorbing medium and the
$K\alpha$ line produces a more accurate representation of the true
\ovi\ column density; in the following analysis we consider both
possibilities.

\subsection{Temperature and Density Constraints}
The abundance ratios of metal ions (for example, \ovii/\oviii) are
expected to vary with temperature as a result of collisional ionization;
additionally, as the density decreases to typical WHIM values
($\sim 10^{-4}$\,cm$^{-3}$), photoionization from the extragalactic
ionizing background plays an increasingly important role.  We used Cloudy
version 90.04 \citep{ferland96} to calculate relative ionic abundances
for all measured elements over a grid of $\log T$ and $\log n_e$.
As with the
Doppler parameter and column density diagnostics described above,
any measured ion column density ratio then produces a ``track'' of consistency
in the $\log T - \log n_e$ plane, and (assuming the ions arise in the 
same gaseous phase) the overlap between such ``tracks'' can place constraints
on the gas temperature and density.

\begin{figure}
\centering
\includegraphics[width=0.8\linewidth]{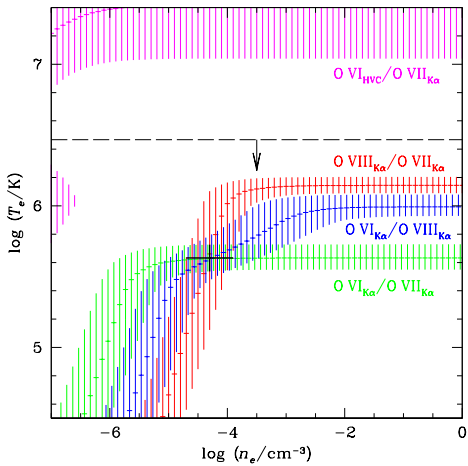}
\caption{$2\sigma$ constraints on the gas temperature and density toward 
Mrk~421,
derived from oxygen ion ratios as labeled.  The dashed line indicates the
upper limit on the temperature from the \ovii\ Doppler parameter.
\label{fig_mkn421_tne}}
\end{figure}

Column density ratios between different ions of the same element 
(e.g. \ovi, \ovii, and \oviii) produce the
strongest constraints since these ratios are independent of the relative
metal abundance in the gas  Figure~\ref{fig_mkn421_tne} shows such constraints
derived from these oxygen abundance ratios.  If the \ovi\ K$\alpha$ line
accurately traces the \ovi\ column density of this medium, then the
contours overlap at $n_e\sim 10^{-4}$\,cm$^{-3}$. Coupled with
the total oxygen column density and assuming O/H of 0.3 times Solar, 
this implies that the absorber has a radial extent of $\sim 1$\,Mpc
and a mass consistent with the expected baryonic mass of the Local 
Group.  

On the other hand, if the \ovi\ K$\alpha$ line does \emph{not} correctly
measure $\novi$, then this absorption can in principle arise in a 
higher--density Galactic medium with $\log T\sim 6.1$ (as derived from
the \oviii/\ovii\ ratio with collisional ionization as the dominant 
process; Figure~\ref{fig_mkn421_tne}).  Even in this case, a low--density
(extragalactic) medium is fully consistent with the data.  Furthermore,
the contour derived from the measured Ne/O abundance (not shown in the
figure) is consistent in the low--density regime if the Ne/O ratio
is significantly supersolar, as has been observed in other 
Galactic and extragalactic absorption systems.

\section{The Mrk~279 Sightline}
A full discussion of our analysis of the Mrk~279 \chandra\ and \fuse\
spectra will appear in a forthcoming paper \citep{williams06}.

\subsection{Observations and Measurements}
While Mrk~421 was observed for relatively short periods during bright
outburst phases, few other bright sources flare this dramatically.
We thus searched the \chandra\ archive for long--duration LETG observations
of relatively bright background quasars.  One such source, Mrk~279
(an AGN at $z=0.03$) was observed during seven periods in May 2003
for a total exposure time of 340 ks.
These seven observations were coadded for a final (unbinned)
signal--to--noise ratio of ${\rm S/N}\sim 6.5$ near 22\AA.
As with the Mrk~421 spectrum, we again used Sherpa to fit a power law and
foreground Galactic absorption to the spectrum over $10-100$~\AA\ band 
(excluding the $49-57.5$~\AA\ and $60.5-67.5$~\AA\ chip gap regions), leaving
the relative Galactic metal abundances 
as free parameters in order to produce a better fit around the absorption
edges.  The remaining broad residuals were
were corrected by including four broad Gaussians in the source model. 

\begin{figure}
\centering
\includegraphics[width=0.8\linewidth]{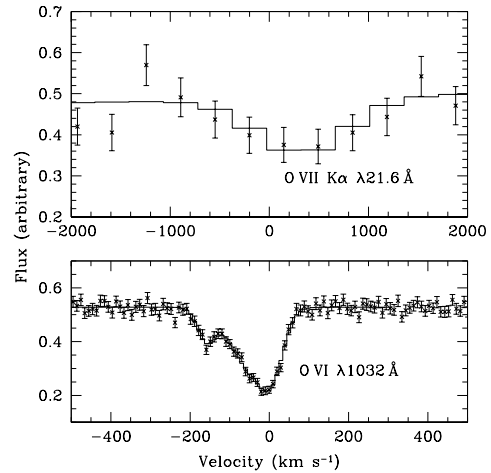}
\caption{Velocity plot of the \ovii\ and \ovi\ absorption seen in the
Mrk~279 \chandra\ and \fuse\ spectra respectively.  The centroids of the
\ovii\ and high--velocity \ovi\ lines differ by $\sim 2.5\sigma$.
\label{fig_mkn279_spec}}
\end{figure}

Although several strong lines such as C\,{\sc vi}, \ovii, and \nvii\ 
are apparent at the blazar redshift ($z=0.03$), only \ovii~K$\alpha$ 
$\lambda 21.602$ is 
unambiguously detected at $21.623\pm 0.012$\AA\ ($v=290\pm 170$\kms) 
with an equivalent width of 
$25.7\pm 5.3$~m\AA.  Upper limits are measured for the \ovii~K$\beta$ 
and \oviii\ lines.
The \chandra\ LETG wavelength scale contains intrinsic random errors 
of approximately
$0.01$\AA, on the order of the statistical error on the \ovii\ position 
measurement, but these errors should not vary with time (J.~J.~Drake, private
communication).  To check the absolute wavelength
scale near the \ovii\ line, we retrieved the nearest HRC--S/LETG calibration 
observation of the X-ray bright star Capella (observation 3675, taken on 2003
September 28) from the \chandra\ archive and 
reprocessed the data in exactly the same manner as the Mrk 279 data.
The wavelength of the strong \ovii\ emission line was found to be 
$21.606\pm 0.002$\AA\ or $56\pm 28$\kms, which is consistent with the 
$+30$\kms\ radial velocity of Capella as listed in the
SIMBAD database.
Thus, any systematic effects on the measured velocity of the \ovii\ absorption
are likely to be insignificant compared to the statistical error.

Mrk~279 was also observed with \fuse\ for a total exposure time 
of 224~ks (though only 177~ks of these data were usable); all calibrated 
data from these observations were retrieved and coadded in the same manner 
as the Mrk~421 \fuse\ data.  This spectrum also shows strong Galactic
thick--disk \ovi\ absorption at $v\sim 0$, but unlike the Mrk~421 spectrum,
the high--velocity \ovi\ is strong and clearly separated from the 
thick--disk \ovi\ at $v= -160\pm 2.6$\kms.  Figure~\ref{fig_mkn279_spec} shows
the \ovii\ and \ovi\ absorption systems plotted against velocity.
Although the error on the \ovii\ velocity is large, it is nonetheless
significantly higher than the \ovi\ HVC velocity by about $2.5\sigma$.
Thus, a direct association between the \ovi\ HVC and \ovii\ can
be ruled out with reasonably high confidence.

\subsection{Doppler parameters}
\begin{figure}
\centering
\includegraphics[width=0.8\linewidth]{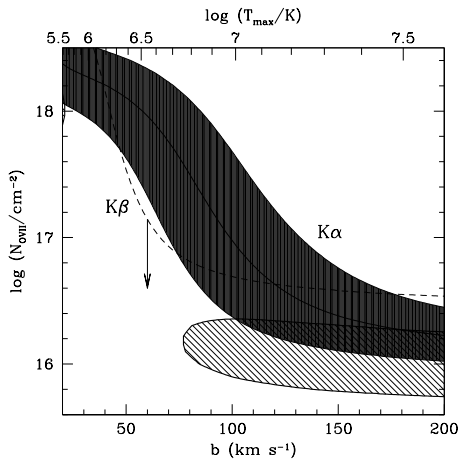}
\caption{Column density and velocity dispersion diagnostics for the Mrk~279
\ovii\ absorption.  The shaded region is derived from the $1\sigma$  
\ovii\ K$\alpha$ equivalent width limits, the dashed line is from the \ovii\ 
K$\beta$ $2\sigma$
upper limit, and the 95\% joint confidence interval is depicted by the
dark hatched region.
\label{fig_mkn279_nhb}}
\end{figure}

In the Mrk~421 \chandra\ spectrum, three absorption lines from the \ovii\ 
K-series were strongly detected.  While a similar curve--of--growth analysis 
can be performed for Mrk~279, placing limits on $\novii$ and $b$ is more
difficult because only the \ovii\ K$\alpha$ line is strongly detected;
only an upper limit can be measured for the K$\beta$ line.
Since the absorption line properties for various column densities
and Doppler parameters are known, limits on these quantities 
can be determined using the \chandra\ spectrum itself.  For each point
in the $\novii-b$ plane, \ovii\ K$\alpha$ and K$\beta$ absorption
lines with the calculated $W_\lambda$ and FWHM values were added to 
the best--fit continuum model, and the $\chi^2$ statistic calculated using
the ``goodness'' command in Sherpa.  Since best--fit K$\beta$ line amplitude
is zero, the unsaturated case (assuming the \ovii\ line ratio constraint 
$W_\lambda(K\beta)=0.15*W_\lambda(K\alpha)$) produces the best fit to the
data.  The minimum $\chi^2$ value was taken from such a fit, and 
$\Delta\chi^2=\chi^2(\novii,b)-\chi_{\rm min}^2$ calculated for every point.  
The 95\% confidence interval ($\Delta\chi^2 < 6$) calculated with this 
method is shown in
Figure~\ref{fig_mkn279_nhb} (as well as the $\novii-b$ contours derived
from each transition); at this confidence
level all Doppler parameters between $20<b<77$\kms\ are ruled out.

Doppler parameters for the \ovi\ absorption were calculated in the same
manner as for Mrk~421.  In this case, however, two Gaussians are necessary
to fit the low--velocity (thick--disk) \ovi.  The derived velocity
dispersions are $b=61.5\pm 3.5$, $38.8\pm 2.8$, and $32.0\pm 4.6$ for the
broad Galactic, narrow Galactic, and high--velocity \ovi\ components
respectively (producing the best--fit model shown in 
Figure~\ref{fig_mkn279_spec}).  These are all strongly inconsistent with the
limits found for the \ovii\ velocity dispersion, indicating that the
\ovii\ is not related to any of the \ovi\ components.  Even if the
low--velocity \ovi\ is considered to be one non--Gaussian component
and its width is measured directly from the spectrum, its Doppler
parameter is $b\sim 80$\kms, barely consistent with the 95\% lower
limit on the \ovii\ $b$ value.  

\begin{figure}
\centering
\includegraphics[width=0.8\linewidth]{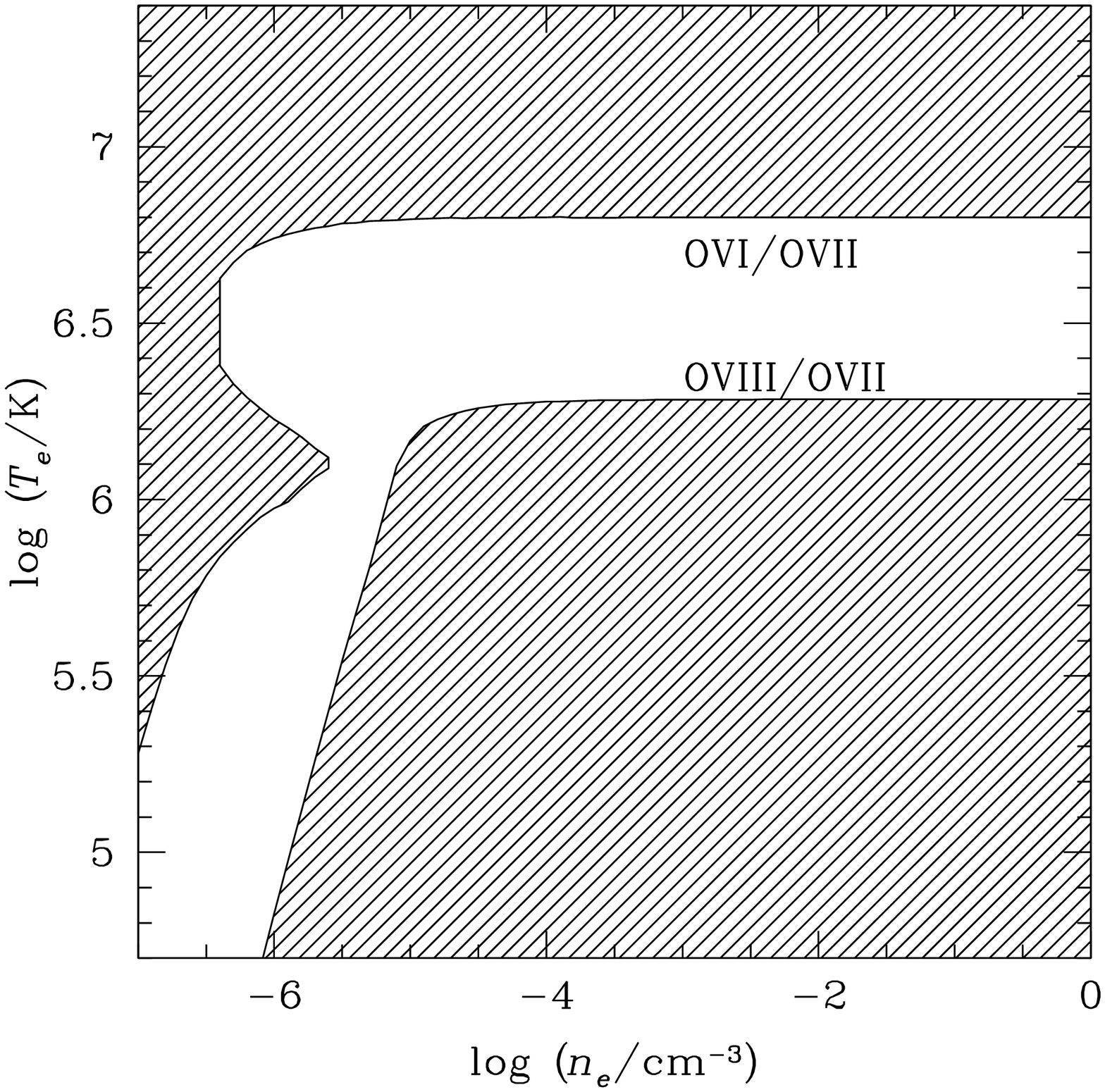}
\includegraphics[width=0.8\linewidth]{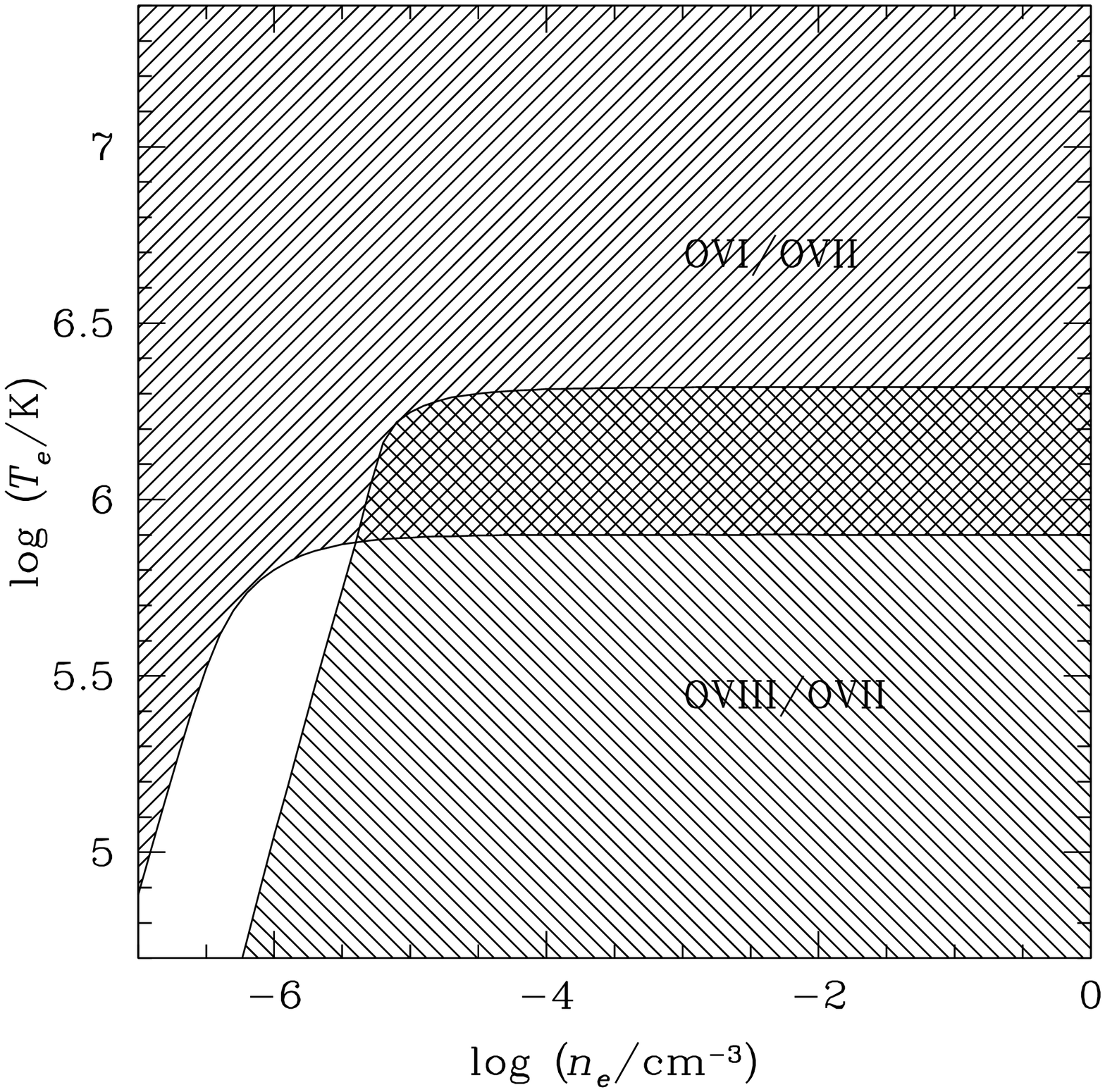}
\caption{Mrk~279 temperature and density constraints from the \oviii/\ovii\ 
and \ovi/\ovii\ upper limits, assuming a putative \ovi\ absorption
line with $b=100$\kms (top) and $b=200$\kms (bottom).
\label{fig_mkn279_tne}}
\end{figure}

\subsection{Temperature and Density Constraints}
An upper limit on the temperature of the absorbing medium of $\log T < 6.3$
can be derived from the \oviii/\ovii\ column density.  Since the
\ovii\ absorption does not appear to be associated with any of the
\ovi\ components, finding a lower limit on the temperature is more
difficult.  We thus assume that the \ovi\ absorption from the
\ovii --bearing gas is undetected in the \fuse\ spectrum, and exists
as a very broad ($b>80$\kms) absorption line superposed on the narrower
detected components.  Limits on the \ovi\ column density associated with
the \ovii\ were thus calculated by placing such an absorption line in the
\fuse\ spectrum model, one with $b=100$\kms\ and one with $b=200$\kms,
and calculating the corresponding \ovi/\ovii\ $T-n_e$ constraint
(shown in Figure~\ref{fig_mkn279_tne}.

The lower limit on temperature derived from the \ovi/\ovii\ upper limit
is highly dependent on $b$; in fact, for $b=100$\kms\ the two oxygen
line ratios are inconsistent with each other for \emph{all} temperatures
and densities.  Thus, if the \ovii\ is associated with an undetected
broad \ovi\ line, the velocity dispersion of the absorption must
be very high (at least $b\sim 200$\kms).  Such a high velocity dispersion,
if purely thermal, implies temperatures of $T\sim 10^7$\,K, which is 
ruled out by the non--detection of \oviii\ absorption.  This absorber
must therefore be broadened primarily by nonthermal processes, perhaps
due to the velocity shear of infalling hot gas associated with HVC
Complex C.

\section{Conclusions}
We have detected strong $z=0$ X-ray absorption toward both Mrk~421 and
Mrk~279.  Both absorption systems appear to exhibit similar column
densities and consistent temperature and density limits (albeit with
large errors).  Additionally, in both cases the detected X-ray absorption
does \emph{not} appear to arise in the same phase as the low-- or 
high--velocity \ovi\ absorption seen in the \fuse\ spectra, indicating
that the \ovii\ absorption likely comes from either the local WHIM or
a heretofore--undiscovered hot Galactic component.  Additionally, the
Doppler parameters of the Mrk~421 and Mrk~279 absorption are highly
inconsistent with each other, perhaps indicating that these two systems
originate from entirely different physical processes.  

\section*{Acknowledgments}
We thank the conference organizers for an enjoyable and informative
week, and the \chandra\ and \fuse\ teams for their exceptional efforts on these
missions.  This work has been supported by \chandra\ award AR5-6017X issued
by the \chandra\ X-ray Observatory Center, which is operated by the
Smithsonian Astrophysical Observatory for and on behalf of the NASA 
under contract NAS8-39073.  RJW derives additional support from an Ohio
State University Presidential Fellowship.

\end{document}